\documentstyle[PASJadd]{PASJ95}
\begin{document}
\setcounter{page}{265}

\title{The Radio and Gamma-Ray Luminosities of Blazars }
\author{L. Zhang,$^{1,2}$  K.S. Cheng,$^2$ J.H. Fan$^{3}$
\\[12pt]
$^1${\it Department of Physics, Yunnan University, Kunming, P.R. China}
\\
$^2${\it Department of Physics, the University of Hong Kong, Hong Kong,
P.R. China}
\\
$^3${\it Center for Astrophysics, Guangzhou University, Guangzhou,
510400, P.R. China}
}

\abst{ 
Based on the $\gamma$-ray data of blazars in the
third EGRET catalog and radio data at 5 GHz, we studied the correlation
between the radio and $\gamma$-ray luminosities using two statistical
methods. The first method was the partial correlation analysis method,
which indicates that there exist correlations between the radio and
$\gamma$-ray luminosities in both high and low states as well as in the
average case. The second method involved a comparison of expected
$\gamma$-ray luminosity distribution with the observed data using
the Kolmogorov--Smirnov (KS) test. In the second method, we assumed that
there is a correlation between the radio and $\gamma$-ray luminosities and
that the $\gamma$-ray luminosity function is
proportional to the radio luminosity function. The KS test indicates that
the expected gamma-ray luminosity distributions are consistent with the
observed data in a reasonable parameter range. Finally, we used different
$\gamma$-ray luminosity functions to estimate the possible 'observed' 
$\gamma$-ray luminosity distributions by GLAST.} 

\kword{galaxies: active --- quasars: general --- gamma rays }

\maketitle

\section{Introduction}

Although the correlation between the radio and $\gamma$-ray luminosities
of blazars has been widely studied, two opposite conclusions have been
made. Based on the data observed by EGRET at different observation
times, a moderate correlation between the radio and $\gamma$-ray
luminosities has been obtained (e.g. Stecker et al. 1993; Padovani et
al. 1993; Salamon, Stecker 1994; Dondi, Ghisellini, 1995; Stecker, Salmon 
, 1996; M\"{u}cke et al. 1997; Mattox et al. 1997; Zhou et al. 1997; Fan
et al. 1998; Cheng et al. 2000). However, M\"{u}cke et al. (1997)
performed Monte Carlo
simulations of selected data sets and applied
different correlation techniques in view of a truncation bias caused by
sensitivity limits of the surveys. In their analysis, they used radio and
$\gamma$-ray luminosity functions given by Dunlop and Peacock (1990) and 
Chiang et al. (1995). After a careful analysis, they claimed that there
is no correlation between the radio and $\gamma$-ray luminosities for
simultaneously observed radio and $\gamma$-ray data. 

Many models have
been proposed to explain the origin of the blazar $\gamma $-ray
emission, including synchrotron self-Compton (e.g.
Maraschi et al 1992), inverse Compton
scattering on photons produced by an accretion disk
(e.g. Dermer et al. 1992;
Zhang, Cheng 1997), or scattered by ambient material, or
reprocessed by broad-line clouds (
e.g. Sikora et al. 1994; Blandford, Levinson 1995; Xie et al. 1997),
synchrotron emission by ultra-relativistic electrons and
positrons (e.g, Ghisellini et al. 1993; Cheng et al. 1993), and
electromagnetic cascade by collision of ultra-relativistic
nucleons (e.g, Mannheim, Biermann 1992; Mannheim 1993;
Cheng, Ding 1994). However, there is no consensus yet
on the dominant emission process.

Correlation analye between different wavebands are important for us
to understand the $\gamma$-ray emission from blazars. For example,
if radio emission is produced by high-energy electrons in the jet
via synchrotron radiation, it is likely that these electrons 
contribute part of the $\gamma$-rays due to inverse Compton 
scattering. In this sense, it seems reasonable to expect some 
correlation between the radio and $\gamma$-ray
luminosities. Here, we revisit the correlation between the radio and
$\gamma$-ray luminosities of blazars using the observed $\gamma$-ray
data in the third catalog of EGRET (Hartman et al. 1999) and radio data
at 5 GHz (see table 1), since the data in these two bands are
presently complete. In section 2, we list the observed data 
and discuss a correlation analysis based on the use of a partial 
correlation analysis. Further, we estimate
$\gamma$-ray luminosity distribution of the observed blazars at 100 MeV
and compared it with the observed data in section 3. Finally, we give our
conclusions and a brief discussion in section 4. 

Throughout the paper we use a Hubble constant, $H_0$, of 75  
km~s$^{-1}$~Mpc$^{-1}$ and the deceleration parameter, $q_0$, of 0.5. 

\section{Observed Data and Correlation Analysis}

The third EGRET catalog includes 66 positive and 25 marginal detections of
$\gamma$-ray loud blazars above 100 MeV (Hartman et al. 1999). We list 67 
of the identified blazars (including Markarian 501) and 25 marginal
blazars in table 1. For each source, its $\gamma$-ray fluxes above 100 MeV
in both the high and low states are the maximum and minimum observed data
in the third EGRET catalog. The average $\gamma$-ray flux is the value
labeled by
P1234 in the third catalog. For radio data at 5 GHz, part of our
sample were taken from those in both high and low states compiled by
Cheng et al. (2000). The average radio data at 5 GHz were taken from
a database provided by the University of Michigan Radio Astronomy
Observatory.  

\begin{table*}
\small
\begin{center}
Table~1. \hspace{4pt}Blazar sample*. 
\end{center}
\vspace{6pt}
\begin{tabular}{lllllllll}
\hline\hline\\
Source$^a$ & z & S$^h_{R}$ (Jy) & S$^l_{R}$ (Jy) 
& $S^{av}_{R}$(Jy)$^b$ & F$^h_{\gamma}$
& F$^l_{\gamma}$ & F$^{av}_{\gamma}$ & $\alpha_{\gamma}$
\\\hline
0202+149 &  1.202&  3.68$\pm$ 0.04&  1.56$\pm$ 0.11&  2.91$\pm$ 0.05&  5.28$\pm$ 2.64&  2.36$\pm$ 0.56&  0.87$\pm$ 
0.28&1.23$\pm$0.15\\
0208-512 &  1.003&  &  3.31$\pm$ 0.12&  & 13.41$\pm$ 2.49&  3.50$\pm$ 1.10&  8.55$\pm$ 
0.45& 0.99$\pm$0.02\\
0219+428 &  0.444&  2.55$\pm$ 0.03&  0.52$\pm$ 0.04&  2.05$\pm$ 0.04&  2.53$\pm$ 0.58&  1.21$\pm$ 0.39&  1.87$\pm$ 
0.29& 1.01$\pm$0.07 \\
0235+164 &  0.940&  4.23$\pm$ 0.02&  0.23$\pm$ 0.05&  1.62$\pm$ 0.06&  6.51$\pm$ 0.88&  1.16$\pm$ 0.40&  2.59$\pm$
0.37& 0.85$\pm$0.06\\
0336-019 &  0.852&  2.29$\pm$ 0.11&  1.18$\pm$ 0.08&  2.39$\pm$ 0.03& 17.76$\pm$ 3.66&  1.31$\pm$ 0.76&  1.51$\pm$
0.35& 0.84$\pm$0.11\\
0420-014 &  0.915&  6.99 &  2.03$\pm$ 0.11&  3.17$\pm$ 0.04&  6.42$\pm$ 3.42&  0.93$\pm$ 0.47&  1.63$\pm$
0.31& 1.44$\pm$0.11\\
0440-003 &  0.844&  1.67$\pm$ 0.08&  1.18$\pm$ 0.02&  1.62$\pm$ 0.07&  8.59$\pm$ 1.20&  2.23$\pm$ 0.41&  1.25$\pm$
0.26& 1.37$\pm$0.10\\
0446+112 &  1.207&  1.89$\pm$ 0.02&  0.50$\pm$ 0.03&  1.55$\pm$ 0.07& 10.90$\pm$ 1.94&  0.63$\pm$ 0.33&  1.49$\pm$
0.25& 1.27$\pm$0.09\\
0454-234 &  1.009&  2.15$\pm$ 0.07&  1.49$\pm$ 0.08&  &  1.47$\pm$ 0.42&  0.81$\pm$ 0.26&  0.81$\pm$
0.26& 2.14$\pm$0.32\\
0454-463 &  0.858&  1.90$\pm$ 0.09&  1.90$\pm$ 0.09&  &  2.28$\pm$ 0.74&  0.55$\pm$ 0.26&  0.77$\pm$
0.21& 1.75$\pm$0.22\\
0458-020 &  2.286&  4.05$\pm$ 0.06&  1.90$\pm$ 0.05&  2.94$\pm$ 0.07&  6.82$\pm$ 4.13&  0.95$\pm$ 0.32&  1.12$\pm$
0.23& 1.45$\pm$0.16\\
0528+134 &  2.070&  6.38$\pm$ 0.10&  1.99$\pm$ 0.04&  3.46$\pm$ 0.09& 35.10$\pm$ 3.68&  3.24$\pm$ 1.43&  9.35$\pm$
0.36& 1.46$\pm$0.02\\
0537-441 &  0.896& 5.30$\pm$ 0.01&  2.52$\pm$ 0.02&  &  9.11$\pm$ 1.46&  1.65$\pm$ 0.45&  2.53$\pm$
0.31& 1.41$\pm$0.07\\
0716+714 &  0.300&  1.17$\pm$ 0.02&  0.31$\pm$ 0.05&  0.62$\pm$ 0.02&  4.57$\pm$ 1.11&  0.93$\pm$ 0.47&  1.78$\pm$
0.20& 1.19$\pm$0.06\\
0735+178 &  0.424&  4.82$\pm$ 0.17&  1.03$\pm$ 0.03&  2.69$\pm$ 0.08&  2.93$\pm$ 0.99&  1.58$\pm$ 0.42&  1.64$\pm$
0.33& 1.60$\pm$0.17\\
0827+243 &  2.050&  0.67&  0.59$\pm$ 0.01&  0.97$\pm$ 0.09& 11.10$\pm$ 6.01&  1.56$\pm$ 0.59&  2.49$\pm$
0.39& 1.42$\pm$0.12\\
0829+046 &  0.180&  2.27$\pm$ 0.20&  0.66$\pm$ 0.18&  1.14$\pm$ 0.03&  3.35$\pm$ 1.63&  1.68$\pm$ 0.51&  1.57$\pm$
0.48& 1.47$\pm$0.24\\
0836+710 &  2.172&  2.70$\pm$ 0.07&  1.65$\pm$ 0.09&  2.13$\pm$ 0.01&  3.34$\pm$ 0.90&  0.86$\pm$ 0.20&  1.02$\pm$
0.47& 1.62$\pm$0.10\\
0851+202 &  0.306&  5.01$\pm$ 0.08&  0.99$\pm$ 0.09&  2.10$\pm$ 0.04&  1.58$\pm$ 0.69&  0.97$\pm$ 0.44&  1.06$\pm$
0.30& 1.03$\pm$0.18\\
0954+556 &  0.901&  2.29$\pm$ 0.11&  1.71$\pm$ 0.20&  1.98$\pm$ 0.01&  4.72$\pm$ 1.55&  0.65$\pm$ 0.25&  0.91$\pm$
0.16& 1.12$\pm$0.10\\
0954+658 &  0.368&  1.62$\pm$ 0.03&  0.18$\pm$ 0.03&  0.64$\pm$ 0.03&  1.80$\pm$ 0.94&  0.66$\pm$ 0.17&  0.60$\pm$
0.15& 1.08$\pm$0.12\\
1101+384 &  0.031&  1.42$\pm$ 0.11&  0.43$\pm$ 0.04&  0.72$\pm$ 0.01&  2.71$\pm$ 0.69&  0.90$\pm$ 0.36&  1.39$\pm$
0.18& 0.57$\pm$0.05\\
1156+295 &  0.729&  2.22$\pm$ 0.04&  0.88$\pm$ 0.02&  1.40$\pm$ 0.03& 16.30$\pm$ 4.00&  0.83$\pm$ 0.20&  0.75$\pm$
0.18& 0.98$\pm$0.11\\
1219+285 &  0.102&  2.39$\pm$ 0.07&  0.54$\pm$ 0.03&  0.94$\pm$ 0.02&  5.36$\pm$ 1.41&  0.69$\pm$ 0.26&  1.15$\pm$
0.78& 0.73$\pm$0.08\\
1222+216 &  0.435&  2.23$\pm$ 0.07&  1.57$\pm$ 0.10&  1.88$\pm$ 0.04&  4.81$\pm$ 1.53&  0.69$\pm$ 0.29&  1.39$\pm$
0.18& 1.28$\pm$0.07\\
1226+023 &  0.158& 44.56$\pm$ 0.38& 30.73$\pm$ 0.89& 37.56$\pm$ 0.22&  4.83$\pm$ 1.18&  0.85$\pm$ 0.42&  1.54$\pm$
0.12& 1.58$\pm$0.06\\
1229-021 &  1.045& 1.04$\pm$ 0.03&  0.94$\pm$ 0.03&  0.94$\pm$ 0.03&  1.55$\pm$ 0.41&  0.49$\pm$ 0.21&  0.69$\pm$
0.15& 1.85$\pm$0.19\\
1253-055 &  0.538& 17.82$\pm$ 0.12& 10.13$\pm$ 0.29& 11.27$\pm$ 0.09& 26.70$\pm$ 1.07&  0.76$\pm$ 0.36&  7.42$\pm$
0.28& 0.96$\pm$0.02\\
1331+170 &  2.084&  &  0.70 &  &  3.31$\pm$ 1.93&  0.94$\pm$ 0.27&  0.44$\pm$
0.16& 1.41$\pm$0.27\\
1406-076 &  1.494&  1.08$\pm$ 0.04&  0.73$\pm$ 0.04&  0.81$\pm$ 0.02& 12.84$\pm$ 2.34&  1.04$\pm$ 0.39&  2.74$\pm$
0.22& 1.29$\pm$0.06\\
1424-418 &  1.522&  4.08$\pm$ 0.14&  2.18$\pm$ 0.09&  &  5.53$\pm$ 1.63&  1.53$\pm$ 0.86&  1.19$\pm$
0.27& 1.13$\pm$0.11\\
1510-089 &  0.361&  4.33$\pm$ 0.04&  0.96$\pm$ 0.14&  2.53$\pm$ 0.04&  4.94$\pm$ 1.83&  1.26$\pm$ 0.53&  1.80$\pm$
0.38& 1.47$\pm$0.12\\
1604+159 &  0.357&  &  0.50&  &  4.20$\pm$ 1.23&  1.23$\pm$ 0.47&  1.28$\pm$
0.41& 1.06$\pm$0.21\\
1606+106 &  1.227&  1.98$\pm$ 0.09&  1.07$\pm$ 0.20&  1.62$\pm$ 0.02&  6.24$\pm$ 1.30&  2.10$\pm$ 0.92&  2.50$\pm$
0.45& 1.63$\pm$0.15\\
1611+343 &  1.404&  4.49$\pm$ 0.03&  2.07$\pm$ 0.09&  3.39$\pm$ 0.05&  6.89$\pm$ 1.53&  1.90$\pm$ 0.40&  2.65$\pm$
0.40& 1.42$\pm$0.09\\
1622-253 &  0.786&  2.34$\pm$ 0.09&  1.53$\pm$ 0.03&  1.94$\pm$ 0.04&  8.25$\pm$ 3.50&  1.01$\pm$ 0.40&  2.12$\pm$
0.35& 1.07$\pm$0.04\\
1622-297 &  0.815&  3.97$\pm$ 0.14&  2.07$\pm$ 0.09&  2.26$\pm$ 0.04& 32.10$\pm$ 3.35&  1.24$\pm$ 0.36&  4.75$\pm$
0.37& 1.07$\pm$0.04\\
1633+382 &  1.814&  3.52$\pm$ 0.03&  1.91$\pm$ 0.02&  2.30$\pm$ 0.01& 10.70$\pm$ 0.96&  3.18$\pm$ 1.04&  5.84$\pm$
0.52& 1.15$\pm$0.05\\
1652+398 &  0.033&  1.96$\pm$ 0.07&  1.04$\pm$ 0.13&  1.46$\pm$ 0.01&  3.20$\pm$ 1.30&  1.80$\pm$ 0.50&  2.50$\pm$
0.70& 0.30$\pm$0.12\\
1730-130 &  0.902&  9.01$\pm$ 0.10&  4.10$\pm$ 0.05&  5.81$\pm$ 0.09& 10.48$\pm$ 3.47&  1.81$\pm$ 0.74&  3.61$\pm$
0.34& 1.23$\pm$0.06\\
1739+522 &  1.375&  3.36$\pm$ 0.03&  0.58$\pm$ 0.05&  1.91$\pm$ 0.06&  4.49$\pm$ 2.69&  0.97$\pm$ 0.47&  1.82$\pm$
0.30& 1.42$\pm$0.13\\
1741-038 &  1.054&  4.90$\pm$ 0.06&  1.50$\pm$ 0.09&  2.88$\pm$ 0.09&  4.87$\pm$ 1.96&  1.76$\pm$ 0.47&  1.17$\pm$
0.33& 1.42$\pm$0.25\\
1830-210 &  1.000&  &  7.92&  &  9.93$\pm$ 2.48&  1.78$\pm$ 0.88& 2.66$\pm$
0.37& 1.59$\pm$0.08\\
1933-400 &  0.966&  1.48$\pm$ 0.08&  0.66$\pm$ 0.01&  &  9.39$\pm$ 3.14& 1.40$\pm$ 0.34&  0.85$\pm$
0.27& 1.86$\pm$0.26\\
2032+107 &  0.601&  1.08$\pm$ 0.04&  0.26$\pm$ 0.17&  0.62$\pm$ 0.02& 3.59$\pm$ 1.50&  1.02$\pm$ 0.38&  1.33$\pm$
0.31& 1.83$\pm$0.17\\
2052-474 &  1.489&  &  2.52$\pm$ 0.10&  &  3.50$\pm$ 2.09&  1.13$\pm$ 0.35&  0.96$\pm$
0.32& 1.04$\pm$0.18\\
2155-304 &  0.116&  0.56$\pm$ 0.07&  0.18$\pm$ 0.09&  0.37$\pm$ 0.02& 3.04$\pm$ 0.77&  0.79$\pm$ 0.35&  1.32$\pm$
0.32& 1.35$\pm$0.15\\
2200+420 &  0.069&  9.98$\pm$ 0.07&  1.69$\pm$ 0.05&  3.10$\pm$ 0.05& 3.99$\pm$ 1.16&  0.88$\pm$ 0.38&  1.11$\pm$
0.31& 1.60$\pm$0.17\\
2209+236 &  1.000&  1.21$\pm$ 0.16&  0.60$\pm$ 0.13&  &  4.57$\pm$ 2.05& 1.23$\pm$ 0.35&  0.69$\pm$
0.23& 1.48$\pm$0.30\\
2230+114 &  1.037&  5.39$\pm$ 0.14&  3.44$\pm$ 0.08&  4.16$\pm$ 0.12& 5.16$\pm$ 1.50&  1.21$\pm$ 0.35&  1.92$\pm$
0.28& 1.45$\pm$0.08\\
2251+158 &  0.859& 24.00$\pm$ 0.72&  7.92$\pm$ 0.14& 10.67$\pm$ 0.14& 11.61$\pm$ 1.84&  2.46$\pm$ 0.96&  5.37$\pm$
0.40& 1.21$\pm$0.03\\
2356+196 &  1.066&  0.86$\pm$ 0.08&  0.59$\pm$ 0.10&  0.73$\pm$ 0.05&  2.63$\pm$ 0.90&  1.28$\pm$ 0.55&  0.83$\pm$
0.28& 1.09$\pm$0.18\\
0414-189$^{\star}$ &  1.536&  &  1.35$\pm$ 0.08&  &  4.95$\pm$ 1.61&  1.37$\pm$ 0.77&  0.91& 2.25$\pm$0.47\\
0430+2859$^{\star}$&  &  &  &  &  7.58$\pm$ 2.21&  1.60$\pm$ 0.33&  2.20$\pm$
0.28& 0.90$\pm$0.05\\
\hline
\end{tabular}

\end{table*}


\begin{table*}
\small
\begin{center}
Table~1. \hspace{4pt} Continued
\end{center}
\vspace{6pt}
\begin{tabular}{lllllllll}
\hline\hline\\
Source$^a$ & z & S$^h_{R}$ (Jy) & S$^l_{R}$ (Jy)
& $S^{av}_{R}$(Jy)$^b$ & F$^h_{\gamma}$
& F$^l_{\gamma}$ & F$^{av}_{\gamma}$ & $\alpha_{\gamma}$
\\\hline
0459+060$^{\star}$ &  1.106&  &  &  &  3.40$\pm$ 1.82&  1.19$\pm$ 0.32&  0.61$\pm$
0.21& 1.36$\pm$0.23\\
0616-116$^{\star}$ &  &  &  &  &  4.79$\pm$ 1.63&  1.49$\pm$ 0.82&  1.09$\pm$
0.34& 1.67$\pm$0.27\\
0738+5451$^{\star}$&  0.723&  &  &  &  4.21$\pm$ 0.83&  1.14$\pm$ 0.63&  1.11$\pm$
0.24& 1.03$\pm$0.10\\
0850-1213$^{\star}$&  0.566&  &  &  &  4.44$\pm$ 1.16&  1.40$\pm$ 0.44&   & 0.58$\pm$0.21\\
1243-072$^{\star}$ &  1.286&  &  1.14$\pm$ 0.04&  &  4.41$\pm$ 2.96&  0.60$\pm$ 0.25&  0.98$\pm$
0.21& 1.73$\pm$0.11\\
1334-127$^{\star}$ &  0.539&  &  &  &  2.02$\pm$ 1.16&  1.14$\pm$ 0.38&  0.55$\pm$
0.19& 1.62$\pm$0.26\\
1725+044$^{\star}$ &  0.296&  &  1.25$\pm$ 0.04&  &  3.02$\pm$ 1.88&  1.33$\pm$ 0.61&  1.79$\pm$
0.41& 1.67$\pm$0.16\\
1759-396$^{\star}$ &  &  &  &  2.29$\pm$ 0.06& 14.57$\pm$ 4.89&  1.75$\pm$ 0.47&  0.98$\pm$
0.29& 2.10$\pm$0.24\\
1908-201$^{\star}$ &  &  2.37$\pm$ 0.03&  2.10$\pm$ 0.05&  0.65$\pm$ 0.03&  3.71$\pm$ 2.03&  1.49$\pm$ 0.34&
1.75$\pm$0.27& 1.39$\pm$0.10\\
1936-155$^{\star}$ &  1.657&  1.69$\pm$ 0.08&  0.63$\pm$ 0.03&  &  5.50$\pm$ 1.86&  0.76$\pm$ 0.30&  0.74 & 2.45$\pm$0.90\\
2022-077$^{\star}$ &  1.388&  &  &  &  7.45$\pm$ 1.34&  2.18$\pm$ 0.38&  2.12$\pm$
0.35& 1.38$\pm$0.10\\
2320-035$^{\star}$ &  1.411&  &  &  &  3.82$\pm$ 1.01&  0.82$\pm$ 0.44&  0.60& \\
2351-456$^{\star}$ &  1.992&  &  &  &  4.28$\pm$ 2.03&  1.18$\pm$ 0.52&  1.43$\pm$
0.37& 1.38$\pm$0.22\\
0234+285 &  1.213&  4.87$\pm$ 0.06&  1.45$\pm$ 0.04&  2.64$\pm$ 0.07&  3.41$\pm$ 1.16&  1.09$\pm$ 0.44&  1.38$\pm$
0.26& 1.53$\pm$0.13\\
0506-612 &  1.093&  1.50$\pm$ 0.08&  1.50$\pm$ 0.08&  &  2.88$\pm$ 1.15&  0.64$\pm$ 0.12&  0.72$\pm$
0.17& 1.40$\pm$0.15\\
0521-365 &  0.055&  8.87$\pm$ 0.17&  6.52$\pm$ 0.16&  8.06$\pm$ 0.04&  3.19$\pm$ 0.72&  1.95$\pm$ 0.44&  1.58$\pm$
0.35& 1.63$\pm$0.26\\
0804+499 &  1.433&  1.94$\pm$ 0.13&  0.22$\pm$ 0.04&  &  1.51$\pm$ 0.61&  0.83$\pm$ 0.39&  1.07$\pm$
0.25& 1.15$\pm$0.24\\
0917+449 &  2.180&  2.34$\pm$ 0.06&  1.00$\pm$ 0.02&  &  3.35$\pm$ 1.30&  1.14$\pm$ 0.33&  1.38$\pm$
0.20& 1.19$\pm$0.08\\
1127-145 &  1.187&  5.62$\pm$ 0.17&  2.95$\pm$ 0.10&  3.63$\pm$ 0.02&  6.18$\pm$ 1.80&  1.08$\pm$ 0.59&  0.99$\pm$
0.24& 1.70$\pm$0.20\\
1313-333 &  1.210&  1.47$\pm$ 0.03&  1.21$\pm$ 0.18&  &  3.18$\pm$ 1.90&  1.62$\pm$ 0.53&  1.46$\pm$
0.25& 1.28$\pm$0.11\\
0119+041$^{\star}$ &  0.637&  2.07$\pm$ 0.09&  1.22$\pm$ 0.06&  &  2.03$\pm$ 0.58&  1.34$\pm$ 0.43&  0.51$\pm$
0.27& 1.63$\pm$0.41\\
0130-171$^{\star}$ &  1.022&  &  1.00$\pm$ 0.07&  &  1.38$\pm$ 0.68&  0.92$\pm$ 0.52&  1.16$\pm$
0.30& 1.50$\pm$0.17\\
0415+379$^{\star}$ &  0.049&  8.67$\pm$ 0.19&  5.81$\pm$ 0.04&  6.44$\pm$ 0.09&  6.02$\pm$ 1.71&  1.02$\pm$ 0.31&  1.28$\pm$
0.26& 1.59$\pm$0.20\\
0537-286$^{\star}$ &  3.110&  1.19$\pm$ 0.10&  1.02$\pm$ 0.07&  1.16$\pm$ 0.08&  3.50$\pm$ 1.18&  0.96$\pm$ 0.49&  0.69$\pm$
0.27& 1.47$\pm$0.36\\
0539-057$^{\star}$ &  0.839&  &  1.56$\pm$ 0.08&  &  &  6.65$\pm$ 1.95&  1.00 & \\
0803+5126$^{\star}$&  1.140&  &  &   &  2.34$\pm$ 1.37&  0.99$\pm$ 0.26&  0.87$\pm$
0.24& 1.76$\pm$0.22\\
1011+496$^{\star}$ &  0.200&  &  &  &  0.80$\pm$ 0.30&  0.44$\pm$ 0.24&  0.48$\pm$
0.14& 0.90$\pm$0.18\\
1055+567$^{\star}$ &  0.410&  &  &  &  1.61$\pm$ 1.01&  0.65$\pm$ 0.16&  0.50$\pm$
0.14& 1.51$\pm$0.28\\
1237+0459$^{\star}$&  &  &  &  1.52$\pm$ 0.86&  0.62$\pm$ 0.21&  0.65$\pm$
0.15& 1.48$\pm$0.27\\
1324+224$^{\star}$ &  1.400&  1.66$\pm$ 0.17&  1.29$\pm$ 0.15&  1.52$\pm$ 0.07&  6.84$\pm$ 2.26&  0.95$\pm$ 0.27&  0.52$\pm$
0.16& 1.58$\pm$0.16\\
1504-166$^{\star}$ &  0.876&  2.72$\pm$ 0.07&  2.12$\pm$ 0.08&  2.39$\pm$ 0.03&  3.32$\pm$ 1.03&  1.65$\pm$ 0.63&
0.88 & \\
1514-241$^{\star}$ &  0.042&  3.00$\pm$ 0.05&  1.35$\pm$ 0.04&  2.23$\pm$ 0.08&  3.72$\pm$ 1.83&  1.85$\pm$ 0.61&  0.84$\pm$
0.28& 1.66$\pm$0.27\\
1716-771$^{\star}$ &  &  &  &  & 4.05$\pm$ 2.31&  1.54$\pm$ 0.57&  0.84$\pm$
0.40& 1.74$\pm$0.24\\
1808-5011$^{\star}$&  &  &  & & 6.21$\pm$ 1.97&  0.73$\pm$ 0.33&  0.59$\pm$
0.27& 1.93$\pm$0.28\\
2105+598$^{\star}$ &  &  &  & & 3.32$\pm$ 0.96&  1.62$\pm$ 0.76&  1.98$\pm$
0.41& 1.21$\pm$0.14\\
2206+650$^{\star}$ &  &  &  & & 3.08$\pm$ 1.32&  1.86$\pm$ 0.82&  2.44$\pm$
0.55& 1.29$\pm$0.15\\
2250+1926$^{\star}$&  &  &  & & 6.22$\pm$ 2.15&  1.06$\pm$ 0.38&  0.58$\pm$
0.28& 1.36$\pm$0.35\\
2346+385$^{\star}$ &  1.032&  &  &  &  3.75$\pm$ 1.03&  0.85$\pm$ 0.36&  0.61$\pm$
0.32& 1.47$\pm$0.40\\

\hline
\end{tabular}\\
*: $\gamma$-ray flux is in units of$10^{-7}$ cm$^{-2}$~s$^{-1}$\\
a: the data of source names which are unmarked by stars are taken from
those compiled by Cheng et al. (1999), the radio data of the source name
which are marked by stars in high state or
low state are taken from K\"{u}hr et al. (1981). \\
b: the averaged radio data are taken from the 
database provided by the University of Michigan Radio Astronomy
Observatory. 

\end{table*}

\begin{figure}
\vbox to5.6in{\rule{0pt}{5.6in}}
\includegraphics{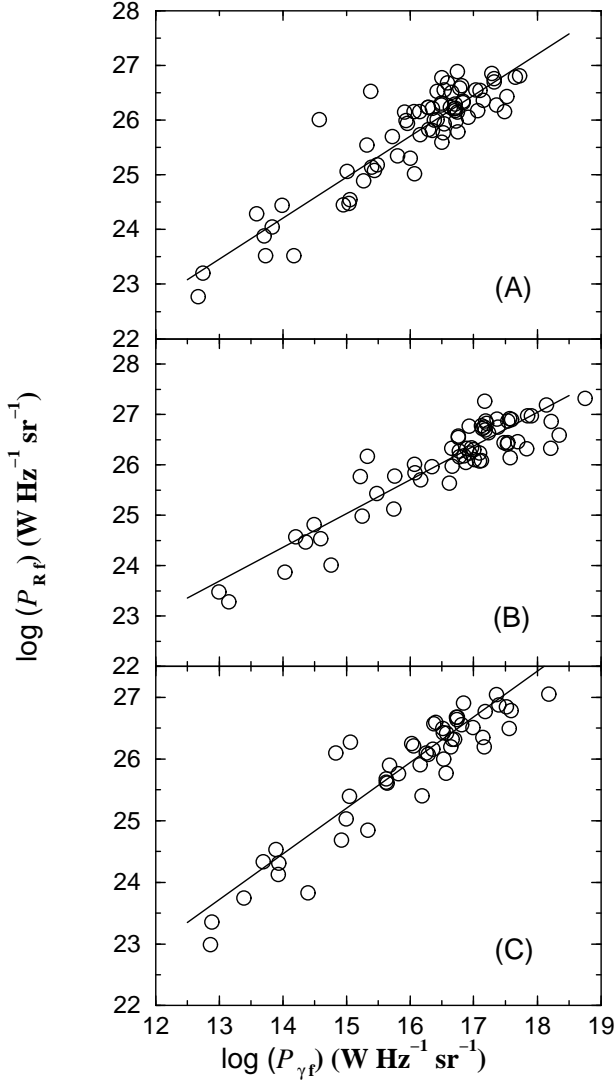}
\caption{Correlations of the $\gamma$-ray luminosity with radio luminosity.
 (A) low state. (B) high state. (C) average case. The differential
luminosity is used in a specific frequencies $\nu$, where $\nu$ is 5 GHz
for the radio and 100 MeV for the $\gamma$-rays.
\label{fm1}}
\end{figure}

From table 1, we can obtain that the mean value of the redshifts in our
sample is $0.96\pm0.06$ and the average value of integrated spectral
index of $\gamma$-rays is $1.22\pm 0.01$ using a weighted average method.
 In order to analyze the correlation between the radio and $\gamma$-ray
luminosities, we made a K-correction for the observed fluxes listed in
table 1 by using
\begin{equation}
F(\nu)=F_{\rm obs}(\nu)(1+z)^{\alpha-1}\;\;\;,
\end{equation}
where $z$ is the redshift and $\alpha$ is the spectral index at the
frequency $\nu$. We used $\alpha=0$ in the radio band and $\alpha=1.22$ in
the $\gamma$ energy range when the spectral indices of the sources were
not known. When the redshift was not known, we used the mean value of our
sample, i.e. $z=0.96$. We now consider the differential luminosity, which
is given by
\begin{equation}
P_{i}(E_i)=P_{i \rm f}\left({E_{i}\over E_{i
\rm f}}\right)^{-\alpha_{i}}\;\;\;,
\label{Pgr}
\end{equation}
where $\alpha_{i}$ is the spectral index, $i={\rm R}$ is for radio and
$i=\gamma$ for $\gamma$-rays. We chose $E_{\gamma \rm f}=100$ MeV. If the
flux density at a
certain energy and the spectral index are known, the differential
luminosity per unit solid angle is given by $P_{i \rm f}=S_{i}(E_{i
\rm f})[R^2_0r^2(1+z)^2]$, where $S_{i}$ is the radio flux density for
$i=\rm R$ or $\gamma$-ray flux density for $i=\gamma$, $z$ is the
redshift, $r$ is the co-moving
coordinate of the source, $R_0$ is the present cosmological scale factor,
and $R_0r=(2c/H_0)[1-(1+z)^{-1/2}]$. Therefore, for radio data,
\begin{eqnarray}
P_{\rm{R f}}=&10^{-30}\left({2c\over
H_0}\right)^2[1-(1+z)^{-1/2}]^2\\\nonumber
&(1+z)^{\alpha_{\rm R}+1}S_{\rm{R~obs}}(E_{\rm{R f}})\;\;\;{\mbox{W~Hz$^{-1}$~sr$^{-1}$}}\;\;,
\label{Pradio}
\end{eqnarray}
where $S_{\rm{R~obs}}(E_{\rm{R f}})$ is in units of Jy. For $\gamma$-ray
data,
\begin{eqnarray}
P_{\gamma \rm f}&\approx 6.6\times 10^{-41}\left({2c\over
H_0}\right)^2[1-(1+z)^{-1/2}]^2\alpha_{\gamma}\\\nonumber
&(1+z)^{\alpha_\gamma+1}F_{\rm{obs}}(>100~{\mbox{MeV}})\;\;{\mbox{W~Hz$^{-1}$~sr$^{-1}$}}\;,
\label{Pgamma}
\end{eqnarray}
where $F_{\rm{obs}}(>~100$MeV) is the integrated $\gamma$-ray flux above
100 MeV in
units of $10^{-7}$ cm$^{-2}$~s$^{-1}$ observed by EGRET and
$\alpha_{\gamma}+1$ is differential spectral index.

Using the observed data in table 1 and equationsq (\ref{Pradio}) and
(\ref{Pgamma}), we obtained the differential luminosities at 5 GHz and 100
MeV. We could thus consider the correlation between the radio and
$\gamma$-ray luminosities in both the high and low states and for
the average case. In the low state, the correlation between luminosities
gives
\begin{equation} 
\rm{log}_{10}P_{\rm{R f}}=(13.7\pm 0.7)+(0.75\pm
0.04)\rm{log}_{10}P_{\rm{\gamma
f}}\;\;\;.
\label{PrPgh}
\end{equation}
 In the high sate, we have 
\begin{equation}
\rm{log_{10}}P_{\rm{R f}}=(15.0\pm 0.7)+(0.67\pm 0.04) \rm{log_{10}}P_{\rm{\gamma f}}
\;\;\;.
\label{PrPgl}
\end{equation}
For the average case, we have
\begin{equation}
\rm{log_{10}}P_{\rm{R f}}=(14.1\pm 0.7)+(0.74\pm 0.04) \rm{log_{10}}P_{\rm{\gamma f}}
\;\;\;.
\label{PrPgav}
\end{equation}
 As pointed out by Padovani (1992), because our sample was flux-limited,
the luminosity was strongly correlated with
redshift, would result in a spurious correlation. Therefore, we
used a partial correlation analysis in our analysis, i.e. we examined the
correlations between the radio and $\gamma$-ray luminosities excluding the
dependence on the redshift. The correlation coefficients used for the
analysis were the Spearman Rank--Order correlation coefficients (Press
et al. 1992). In the low state (72 sources),  the partial correlation
coefficient is $r\approx 0.26$ (with chance probability $P\approx
2.6\times 10^{-2}$), which was derived by
using the fact that the correlation coefficients between log
$P_{\rm{\gamma f}}$ and log $P_{\rm{R f}}$, log $P_{\rm{\gamma f}}$ and log $z$, and
log $P_{\rm{R f}}$ and log $z$ are 0.904, 0.972, and 0.903 respectively. In the high
state (62 sources), the correlation coefficients
between log $P_{\rm{\gamma f}}$ and log $P_{\rm{R f}}$, log $P_{\rm{\gamma f}}$ and log
$z$, and log $P_{\rm{R f}}$ and log $z$ are 0.919, 0.967, and 0.919
respectively; the partial correlation coefficient is thus $r\approx 0.30$
(with chance probability $P\approx 1.8\times 10^{-2}$). 
Therefore, we may conclude that there exist correlations between the radio
and $\gamma$-ray luminosities in both the high and low states. For the average
case, from table 1, there are 51 sources which have average observed radio data. 
 The average $\gamma$-ray flux for each source is the value (labeled
P1234) in the third EGRET catalog (Hartman et al. 1999). The partial
correlation analysis gives $r\approx 0.28$ with a chance probability, 
$P$, of $\approx 3.9\times 10^{-2}$, which is at least a marginal
correlation. Figure \ref{fm1} shows the correlation between the luminosities for
the above three cases.

\section{Gamma-Ray Luminosity Distribution}

\begin{figure}
\vbox to5.0in{\rule{0pt}{5.0in}}
\includegraphics{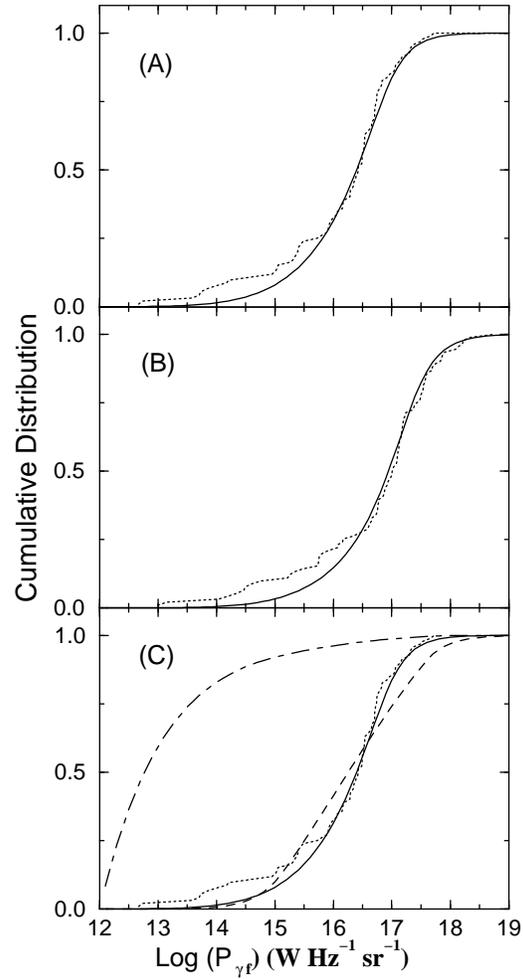}
\caption{The Comparison of observed and expected cumulative distributions
of the blazar $\gamma$-ray luminosities at 100 MeV. (A) low state. (B)
high state. (C) average case. For the average case, the $\gamma$-ray
luminosity distributions calculated by using Eqs. (\ref{nobs_2}) and
(\ref{nobs_3}) are represented by solid, dashed ($\gamma_1=1.2$ and
$\gamma_2=2.2$) and dot-dashed ($\gamma_1=2.9$ and $\gamma_2=2.6$) curves
respectively. Dotted curves represent the observed distributions. 
\label{fm2}}
\end{figure}

We now consider the $\gamma$-ray luminosity distribution in order
to further discuss whether there is a correlation between the radio and
$\gamma$-ray luminosities. Using a preexisting radio luminosity
function and assuming that the radio and $\gamma$-ray luminosity are
related [the correlation coefficients are given by equations (\ref{PrPgl}),
 (\ref{PrPgh}) and (\ref{PrPgav})], we could calculate the $\gamma$-ray
luminosity distributions in the low, high states, and average cases following
the method given by Stecker et al. (1993). As a comparison, when we
calculated the $\gamma$-ray luminosity distribution in the average case, we
used $\gamma$-ray luminosity function given by Chiang and Mukherjee (1998).
 These expected distributions were compared to the
observed distribution. In figure \ref{fm2}, the normalized cumulative
observed $\gamma$-ray luminosity distributions in the low, high states, and the
average case are shown (dotted curves). 

\begin{figure}
\vbox to1.75in{\rule{0pt}{1.75in}}
\includegraphics{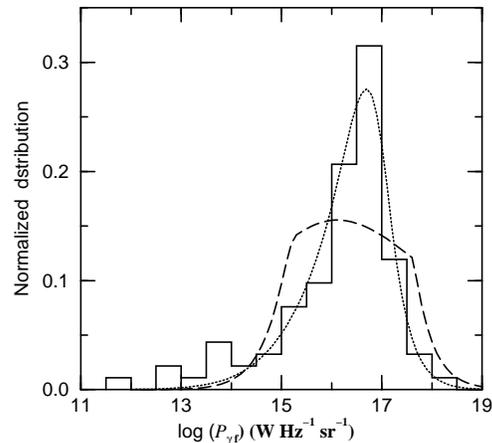}
\caption{The comparison of the differential $\gamma$-ray luminosity
distributions calculated  by using Eqs. (\ref{nobs_2}) (dotted curve) and
(\ref{nobs_3}) (dashed curve) with the observed data (histogram). 
\label{fm3}}
\end{figure}
   
If the $\gamma$-ray luminosity function ($\rho_{\gamma}$) is given, the
total number of sources of the luminosity, $P_{\rm{\gamma f}}$, seen at Earth is
(Salamon, Stecker 1994)
\begin{eqnarray}
n_{\rm{obs}}d({\mbox{log}}_{10} P_{\rm{\gamma f}})&=
\int^{R_0r_{\rm{max}}}_0 4\pi R^3_0r^2\\\nonumber
&\rho_{\gamma}(P_{\rm{\gamma f}},z)
dr~d({\mbox{log}}_{10} P_{\rm{\gamma f}}) \;\;\;,
\label{nobs_1}
\end{eqnarray}
where $P_{\rm{\gamma f}}$ is given by equation (\ref{Pgr}) for a source with
spectral index $\alpha_{\gamma}$, and $R_0r_{\rm{max}}$ is the maximum
detectable distance which is found by using equation (\ref{Pgamma})
and substituting the flux sensitivity, $F_{\rm{\gamma lim}}$, for $F_{\rm{obs}}$. 
 We can convert $R_0r_{\rm{max}}$ to $z_{\rm{max}}$ using the relation of 
$R_0r$ and $z$. In other words, the maximum redshift can be
estimated by
\begin{eqnarray}
&(1+z_{\rm{max}})^{(\alpha_{\gamma}+1)}[(1+z_{\rm{max}})^{-1/2}-1]^2=\\\nonumber
&P_{\rm{\gamma~f}}/[(2c/H_0)^2S_{\rm{\gamma~lim}}]\;\;\;.
\label{zmax}
\end{eqnarray}
It should be noted that the co-moving density, $\rho_{\gamma}$, is 
in units of Mpc$^{-3}\times$(unit interval of log$_{10}P_{\gamma})^{-1}$. 

Salamon and Stecker (1994) assumed that (i) there is a linear
relation between the $\gamma$-ray luminosity ($P_{\rm{\gamma f}}$) and the radio
luminosity, i.e. $P_{\rm{\gamma f}}=10^{\xi}P_{\rm{R f}}$ and (ii)
$\rho_{\gamma}=\eta\rho_{\rm{R}}$, where $\eta$ is a normalization factor and
$\rho_{\rm{R}}$ is a radio luminosity function. They found $\xi=-9.8$ by fitting
38 EGRET sources. Here, we assume that $P_{\rm{R f}}=10^aP^b_{\rm{\gamma f}}$,
where the values of $a$ and $b$ in the high, low states, and for average
case were estimated by equations (\ref{PrPgh}), (\ref{PrPgl}), and
(\ref{PrPgav}) respectively. The radio luminosity function is
given by Dunlop and Peacock (1990)
\begin{equation}
\rho_{\rm R}(P_{\rm R},z)\sim \left\{\left[{P_{R}\over P_c(z)}\right]^{0.83}
+\left[{P_{R}\over P_c(z)}\right]^{1.96}\right\}^{-1}\;\;\;,
\label{rho_r}
\end{equation}
where Log$P_c(z)=25.26+1.18z-0.28z^2$, $P_{\rm R}$ is differential radio
luminosity in units of W~Hz$^{-1}$~sr$^{-1}$, and $\rho_{\rm R}$ is co-moving
density in units of Mpc$^{-3}\times$(unit interval of
log$_{10}P_{\rm R})^{-1}$. Using above assumptions and equation (\ref{rho_r}), 
 we have
\begin{eqnarray}
&n_{\rm {obs}}d({\mbox{log}}_{10} P_{\rm{\gamma f}})=2\pi\eta\left({2c\over
H_0}\right)^3\int^{z_{\rm{max}}}_0\\\nonumber
&{[(1+z)^{-1/2}-1]^2
\over (1+z)^{5/2}} 
\rho_{\rm R}(10^aP^b_{\rm{\gamma f}},z)
dz~d({\mbox{log}}_{10} P_{\rm{\gamma f}}) \;\;\;,
\label{nobs_2}
\end{eqnarray}
where $z_{\rm{max}}$ is determined by equation (\ref{zmax}). Generally, once 
the $z_{\rm{max}}$ is given, we 
can obtain the $\gamma$-ray luminosity distribution. Because we assumed a 
maximum cutoff redshift of $z_{\rm{max}}=5$, the upper limit of 
equation (\ref{nobs_2}) is min ($z_{\rm{max}}$, 5). In order to estimate the 
$\gamma$-ray luminosity distribution using equation (\ref{nobs_2}), we 
needed to know the $\gamma$-ray flux sensitivity and to then determine 
two parameters
($a$, $b$). We adjusted $a$ and $b$ within the allowed range, which is limited
by equation (\ref{PrPgh}) in the low state or by equation (\ref{PrPgl})in the high state.   
 We used $a=13.7$, $b=0.76$, $F_{\rm{\gamma~lim}}(>~100$ MeV)$=0.5\times
10^{-7}$cm$^{-2}$~s$^{-1}$ in the low state, $a=14.0$, $b=0.68$,
$F_{\rm{\gamma~lim}}(>~100$ MeV)$=1.5\times 10^{-7}$ cm$^{-2}$~s$^{-1}$ in the high
state, and $a=14.1$, $b=0.74$ and $F_{\rm{\gamma~lim}}(>~100$ MeV)$ =0.5\times
10^{-7}$ cm$^{-2}$~s$^{-1}$ for the average case. We used one sample KS
test to compare the cumulative distributions with the observed ones. The
maximum deviation are 0.09 in the low state (92 sources), 0.1 in the high state
(91 sources) and 0.07 for the average case (92 sources). Therefore, the
null hypothesis cannot be rejected at the $>80\%$ confidence level. 
 Comparisons of the observed data and our results in the low, high states, and
the average case are shown in panels (A), (B), and (C) of figure \ref{fm2}, 
respectively.
\begin{figure}
\vbox to3.0in{\rule{0pt}{3.0in}}
\includegraphics{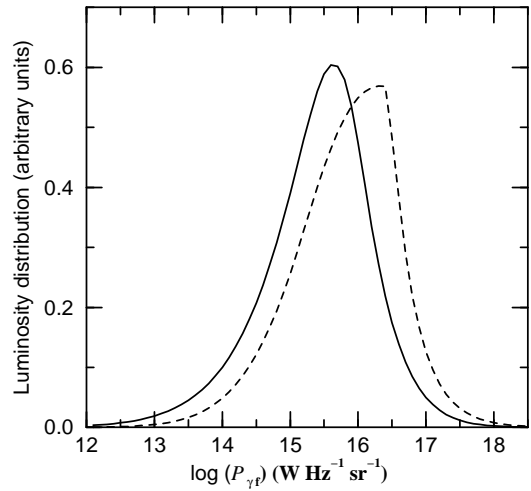}
\caption{The expected differential distributions of average $\gamma$-ray
luminosities of the blazars at 100 MeV. The threshold of GLAST is used
($F_{\gamma~lim}(>100$ MeV) $=3\times 10^{-9}$ cm$^{-2}$ s$^{-1}$). The 
$\gamma$-ray luminosity distributions calculated by using Eq. (\ref{nobs_2}) 
and Eq. (\ref{nobs_3}) are represented by solid and dashed curves respectively. 
\label{fm4}}
\end{figure}

For a comparison, in the average case, we could also estimate the average
$\gamma$-ray luminosity distribution using the $\gamma$-ray luminosity
function of the EGRET blazars.  The $\gamma$-ray luminosity function can
be expressed as (Chiang and Mukherjee 1998)
\begin{eqnarray}
\rho_{\rm{\gamma~EGRET}}&\sim \left({P_{\gamma 0}\over P_B}\right)^{-\gamma_2}
\Theta(P_{\gamma 0}-P_B)+\\\nonumber
&\left({P_{\gamma 0}\over P_B}\right)^{-\gamma_1}
\Theta(P_B-P_{\gamma 0})\;\;\;,
\label{rho_g}
\end{eqnarray}
where $\gamma_1\approx 1.2$, $\gamma_2=2.2\pm 0.1$, log$P_B\approx
14.9$ W~Hz$^{-1}$~sr$^{-1}$ (corresponding to $L_B\approx 1.1\times
10^{46}$ erg~s$^{-1}$) and $\Theta(x)$ is a step function. $P_{\gamma
0}=P_{\gamma}/f(z)$, where $f(z)=(1+z)^{\beta}$ with $\beta=2.7\pm 0.7$.
Using equation (\ref{rho_g}), we can rewrite equation (\ref{nobs_1}) as
\begin{eqnarray}
&n_{\rm{obs}}d({\mbox{log}}_{10} P_{\rm{\gamma f}})=4\pi\rm{log}_{\rm e}10\left({2c\over
H_0}\right)^3\int^{z_{\rm{max}}}_0\\\nonumber
&{[(1+z)^{-1/2}-1]^2\over (1+z)^{5/2}} 
\rho_{\rm{\gamma~EGRET}}P_{\rm{\gamma f}}
dz~d({\mbox{log}}_{10} P_{\rm{\gamma f}}) \;\;
\label{nobs_3}
\end{eqnarray}
where the upper limit of the integral is min($z_{\rm{max}}$, 5). We calculated
the $\gamma$-ray luminosity distribution using equation (\ref{nobs_3}). In
our calculation, we used the minimum value of the average
$\gamma$-ray flux in our sample as the flux sensitivity, which is
consistent with that used by Chiang and Mukherjee (1998), and 
$\beta=2.2$.  The maximum deviation is 0.12 for the expected
$\gamma$-ray luminosity distribution calculated by equation (\ref{nobs_3}).
Therefore, the KS test indicates that the null hypothesis cannot be
rejected at the $>90\%$ confidence level. A comparison of the observed
and expected cumulative $\gamma$-ray luminosity distributions is shown in
the panel (C) of  figure \ref{fm2}.

\section{Conclusion and Discussion}

Using $\gamma$-ray data of the blazars in the third EGRET catalog
(Hartman et al. 1999) and radio data at 5 GHz, we used two methods
to consider the correlation between the radio and $\gamma$-ray luminosities
in the high and low states as well as in the average case. We first made a
partial correlation analysis for the radio and $\gamma$-ray luminosities. Our 
results indicate that there are, at least marginal, correlations between
the radio and $\gamma$-ray luminosities for the three cases. Second, we 
calculated the $\gamma$-ray luminosity distributions of the blazars
by assuming a correlation between the radio and $\gamma$-ray luminosities (the
correlation coefficients were determined by the observed data, see equations (
\ref{PrPgl}), (\ref{PrPgh}) and (\ref{PrPgav}) for both the low and high
states as well as the average case) and using the $\gamma$-ray luminosity
functions given by equation (\ref{nobs_2}). We found that the expected
$\gamma$-ray luminosity distributions are consistent with the observed
data by using KS test (see figure \ref{fm2}). Therefore, after using  two
statistical methods, we conclude that there is at least some correlation
between the radio and $\gamma$-ray luminosities within a reasonable parameter
range. 

We also used the $\gamma$-ray luminosity functions given by Chiang
and Mukherjee (1998) to estimate the $\gamma$-ray luminosity 
distribution. It should be noted that in their calculation of the
$\gamma$-ray luminosity function, Chiang and Mukherjee (1998) used the 1 Jy radio
catalog of K\"{u}hr et al. (1981) as an additional
criterion to reduce the selection effects resulting from an incomplete redshift
sample of AGNs. Therefore, there are 34 AGNs in their sample. The
excluded blazars either fail to meet the significance limit,
 $(TS)^{1/2}=4$, or the 1 Jy radio flux limit. In the calculation, they
introduced a radio completeness function [see equation (7) of  their paper] to
reflect that the sample is restricted to being in the K\"{u}hr catalog.
 Therefore, although they did not explicitly assume a correlation between
the radio and $\gamma$-ray luminosities, they did include an assumption that all
blazars must be observed at the radio band and that their radio energy fluxes must
be greater than 1 Jy. Obviously, this assumption is a key factor by which
the expected gamma-ray luminosity distribution can roughly explain the
observed data. In order to account for it, we note that Chiang et al.
(1995) used the first EGRET catalog to obtain a gamma-ray luminosity function by
modeling the optical evolution of their sources simultaneously with the 
gamma-ray evolution. The obtained $\gamma$-ray luminosity function is the
same as equation (\ref{rho_g}) with different indexes ( $\gamma_1=2.9$ and
$\gamma_2=2.6$). When we used the result of Chiang et al. (1995) to
estimate the
gamma-ray luminosity distribution, the result was not consistent with the
new EGRET catalog at all (see dot-dashed curve in the panel (C) of figure
{\ref{fm2}).

From the KS test, we can see that the $\gamma$-ray luminosity
distributions calculated by using equation (\ref{nobs_2}) is better than that
calculated by using equation (\ref{nobs_3}). The former indicates a peak at
log$P_{\rm{\gamma f}}\sim 16.75$, which is
consistent with the observed data. The later indicates a wider peak from
log$P_{\rm{\gamma f}}\sim 15$ to log$P_{\rm{\gamma f}}\sim 17.5$. In figure \ref{fm3},
we show a comparison of the expected differential $\gamma$-ray
luminosity distributions with the observed data. Further, we use the
possible threshold of GLAST, $F_{\rm{\gamma~lim}}\sim 3\times 10^{-9}$
cm$^{-2}$ s$^{-1}$ (Kamae et al. 1999), to expect the $\gamma$-ray
luminosity distributions using equations (\ref{nobs_2}) and (\ref{nobs_3}),
 respectively. The results are
shown in figure \ref{fm4}. Here, we have normalized the distributions to make
the areas covered by the curves  be the same. It can be seen that they
have roughly the same shapes, but the former peaks are at log$P_{\rm{\gamma
f}}=15.6$ and the later at log$P_{\rm{\gamma f}}=16.3$. We expect that this
difference will be examined by GLAST.

From an observational point of view, there are some differences 
between BL Lacertae objects and flat spectral radio quasars (FSRQs), 
with the former showing no/weak emission lines, steeper X-ray spectra,
 and the later showing strong emission lines and flatter X-ray spectra.
They are discussed separately in some cases (see Fossati et al. 1998;
Ghisellini et al. 1998). 
But they both show a similar flat radio spectrum
and a similar GeV $\gamma$-ray spectrum. Since we only considered the
radio and the $\gamma$-ray regions,  we discussed them both together.

Based on a statistical analysis, we have discussed the
correlation between the $\gamma$-ray
and the radio regions using the higher frequency radio data, the higher
and lower states $\gamma$-ray and the radio regions (Fan et al. 1998; Cheng 
et al. 2000).  We found that there is a closer correlation
between the $\gamma$-ray emission and the high frequency (1.3 mm, 230 GHz)
radio emission for the maximum data than between the $\gamma$-ray and the
lower frequency (5GHz) radio emission (Fan et al. 1998). When we revisited
the multiwavelength correlation using the compiled data, we found that
there are very strong correlations
between $F_{\rm X}$ and $F_{\rm O}$, and $F_{\rm O}$ and $F_{\rm K}$
in both the low and high states. However, a strong correlation
between $F_{\rm X}$ and $F_{\rm K}$ exists only in the low state.
There are also hints of an anti-correlation between $F_{\rm{\gamma}}$ and
$F_{\rm X}$ as well as $F_{rm{\gamma}}$ and $F_{\rm O}$, and a positive correlation
between $F_{\gamma}$ and $F_R$. But the correlation between the
$\gamma$-ray and the radio is not as strong as compared with those
between $F_{\rm X}$ and $F_{\rm O}$, and $F_{\rm O}$ and $F_{\rm K}$ (Cheng et al. 2000).
Our present result is consistent with our previous ones.
Besides, we wish to mention that there are 10  sources without any 
available redshift. There is no difference between the results obtained
with those 10 sources both excluded and included.
 
Physically, if high-energy electrons in the jet are responsible for the
radio and $\gamma$-ray emission, it would then be possible that radio photons
are produced by the synchrotron radiation of these electrons, and that
$\gamma$-rays are produced by the inverse Compton scattering of ambient
lower energy photons, which may be the synchrotron photons. Therefore, 
there should exist a correlation between the radio and $\gamma$-ray luminosity
to some extent. In fact, it is one of the predictions in the synchrotron
self-Compton (SSC) models (e.g. Maraschi et al. 1992).
 Obviously, the seed photons may also be radiated directly from the accretion
disk or disk photons scattered by a broad line region or wind material
above the disk into the jet path (e.g. Dermer, Schlickeiser 1993;
Sikora et al. 1994; Blandford, Levinson 1995; Zhang, Cheng
1997). 

\par
\vspace{1pc} \par
This work is partially supported by the Outstanding Researcher Awards of the
University of Hong Kong, a Croucher Foundation Senior Research
Fellowship and the National 973 projection of China (NKBRSF G19990754).
 This research has made use of data from the University of
Michigan Radio Astronomy Observatory, which is supported by funds from the
University of Michigan. 

\section*{References}
\small

\re
Blandford, R.D., Levinson, A.\ 1995, ApJ, 441, 79
\re 
Cheng, K.S., Ding, W.K.Y.\ 1994, A\&A, 288, 97
\re
Cheng, K.S., Yu, K.N.,\& Ding, K.Y.\ 1993, A\&A, 275, 53
\re
Cheng, K.S., Zhang, X., \& Zhang, L. 2000, ApJ, 537, 80
\re
Chiang, J., Fichtel, C.E., von Montigny, C., Nolan, P.L. \& Petrosian, V. 
\ 1995, ApJ, 452, 156
\re
Chiang, J., \& Mukherjee, R. \ 1998, ApJ, 496, 752
\re
Dermer, C.D., \& Schlickeiser, R. \ 1993, ApJ, 416, 458
\re
Dermer, C.D., Schlickeiser R., \& Mastichiadis A.\ 1992, A\&A,
 256, L27
\re
Dondi, L., \& Ghisellini, G.\ 1995, MNRAS, 273, 583 
\re
Dunlop, J.S., \& Peacock, J.A.\ 1990, MNRAS, 247, 19
\re
Fan, J.H., Adam, G., Xie, G.Z., Cao, S.L., Lin, R.G., \& Copin, Y.\
1998, A\&A, 338, 27
\re
Ghisellini, G., Padovani, P., Celotti, A., Maraschi, L.\ 1993, ApJ, 407,
65
\re
Hartman, R.C. Bertsch, D.L., Bloom, S.D., Chen, A.W., Deines-Jones, P.,
Esposito, J.A., Fichtel, C.E., Friedlander, D.P., et al.\ 1999, ApJS,
123, 79 
\re
Kamae, T., Ohsugi, T., Thompson, D.J., Watanabe, K.\ 1999,
astro-ph/9901187
\re
K\"{u}hr, H., Witzel, A., Pauliny-Toth, I.I.K., Nauber, U. \ 1981, A\&AS,
45, 367
\re
Mannheim, K. \ 1993, A\&A, 269, 67
\re
Mannheim, K., \& Biermann, P.L. \ 1992, A\&A, 253, L21
\re
Maraschi, L., Ghisellini, G., \& Celotti, A. 1992, ApJ, 397, L5
\re
Mattox, J.R., Schachter, J., Molnar, L., Hartman, R.C., \& Patnaik, A.R.
\ 1997, ApJ, 481, 95
M\"{u}cke, A., Pohl, M., Reich,P., Reich, W., Schlickeiser, R., Fichtel,
C.E., Hartman, R.C., Kanbach, G., et al. \ 1997, A\&A, 320, 33
\re
Padovani, P. 1992, A\&A, 256, 399
\re
Padovani, P., Ghisellini, G., Fabian, A.C., \& Celotti, A. \ 1993, MNRAS,
260, L21
\re
Press, W., Flannery, B., Teukolsky, S., Vetterling, W. 1992, Numerical
Recipes: The Art of Scientific Computing 2nd ed., (Cambridge
Cambridge Univ. Press), p
\re
 Salamon, M.H. \& Stecker, F.W. \ 1994, ApJ, 430, L21
\re
Sikora, M., Begelman, M.C., \& Rees, M.J. \ 1994, ApJ, 421, 153
\re
Stecker, F.W., \& Salmon, M.H. \ 1996, ApJ, 464, 600
\re
Stecker, F.W., Salmon, M.H., \& Malkan, M.A. \ 1993, ApJ, 410, L71
\re
Xie, G.Z., Zhang, Y.H., \& Fan, J.H. \ 1997, ApJ, 477, 114
\re
Zhang, L., \& Cheng, K.S. \ 1997, ApJ, 488, 94
\re
Zhou, Y.Y., Lu, Y.J., Wang, T.G., Yu, K.N., Young, E.C.M.\ 1997,
ApJ, 484, L47

\label{last}

\end{document}